# The bromodomain-containing protein Ibd1 links multiple chromatin related protein complexes to highly expressed genes in *Tetrahymena thermophila*.

Alejandro Saettone[1], alejandro.saettone@ryerson.ca

Jyoti Garg[*,2], jyoti@yorku.ca

Jean-Philippe Lambert[*,3,‡], Jean-Philippe.Lambert@crchudequebec.ulaval.ca

Syed Nabeel-Shah[1, Ł], nabeel.haidershah@mail.utoronto.ca

Marcelo Ponce[4], mponce@scinet.utoronto.ca

Alyson Burtch[1], aburtch@ryerson.ca

Cristina Thuppu Mudalige[1], ctthuppu@ryerson.ca

Anne-Claude Gingras[3, 5], gingras@lunenfeld.ca

Ronald E. Pearlman[2], ronp@yorku.ca

Jeffrey Fillingham[1, †], jeffrey.fillingham@ryerson.ca

1. Department of Chemistry and Biology, Ryerson University, 350 Victoria St., Toronto, M5B 2K3, Canada.
2. Department of Biology, York University, 4700 Keele St., Toronto M3J 1P3, Canada.
3. Lunenfeld-Tanenbaum Research Institute at Mount Sinai Hospital, Toronto, M5G 1X5, Canada.
4. SciNet HPC Consortium, University of Toronto, 661 University Ave, Suite 1140, Toronto, M5G 1M1, Canada.
5. Department of Molecular Genetics, University of Toronto, Toronto, M5S 1A8, Canada.

‡. Present address: Department of Molecular Medicine, Université Laval, Quebec, Canada; Centre Hospitalier Universitaire de Québec Research Center, CHUL, 2705 Boulevard Laurier, Quebec, G1V 4G2, Canada.

Ł Present address: Department of Molecular Genetics, University of Toronto, Toronto, M5S 1A8, Canada.

* Equal contribution

† To whom correspondence should be addressed. Tel: +1 (416) 979-5000, Ext: 2123; Email: Jeffrey.fillingham@ryerson.ca

# ABSTRACT

**Background:** The chromatin remodelers of the SWI/SNF family are critical transcriptional regulators. Recognition of lysine acetylation through a bromodomain (BRD) component is key to SWI/SNF function; in most eukaryotes, this function is attributed to SNF2/Brg1.

**Results:** Using affinity purification coupled to mass spectrometry (AP-MS) we identified members of a SWI/SNF complex (SWI/SNF$^{Tt}$) in *Tetrahymena thermophila*. SWI/SNF$^{Tt}$ is composed of 11 proteins, Snf5$^{Tt}$, Swi1$^{Tt}$, Swi3$^{Tt}$, Snf12$^{Tt}$, Brg1$^{Tt}$, two proteins with potential chromatin interacting domains and four proteins without orthologs to SWI/SNF proteins in yeast or mammals. SWI/SNF$^{Tt}$ subunits localize exclusively to the transcriptionally active macronucleus (MAC) during growth and development, consistent with a role in transcription. While *Tetrahymena* Brg1 does not contain a BRD, our AP-MS results identified a BRD-containing SWI/SNF$^{Tt}$ component, Ibd1 that associates with SWI/SNF$^{Tt}$ during growth but not development. AP-MS analysis of epitope-tagged Ibd1 revealed it to be a subunit of several additional protein complexes, including putative SWR$^{Tt}$, and SAGA$^{Tt}$ complexes as well as a putative H3K4-specific histone methyl transferase complex. Recombinant Ibd1 recognizes acetyl-lysine marks on histones correlated with active transcription. Consistent with our AP-MS and histone array data suggesting a role in regulation of gene expression, ChIP-Seq analysis of Ibd1 indicated that it primarily binds near promoters and within gene bodies of highly expressed genes during growth.

**Conclusions:** Our results suggest that through recognizing specific histones marks, Ibd1 targets active chromatin regions of highly expressed genes in *Tetrahymena* where it subsequently might coordinate the recruitment of several chromatin remodeling complexes to regulate the transcriptional landscape of vegetatively growing *Tetrahymena* cells.

**BACKGROUND**

Eukaryotic cells possess multiple levels of regulation of mRNA transcription by RNA polymerase II. Many co-activators of transcription exert their function through chromatin modifying activities. In budding yeast, the SAGA histone acetyl transferase complex co-activates transcription by acetylating specific lysine residues in the N-terminus of histone H3 within the nucleosome, which can then serve as a platform to recruit the SWI/SNF complex via the bromodomain (BRD) present in SNF2/Brg1 [1]. The BRD specifically binds acetyl lysine (Kac) within proteins such as histones [2]. When recruited to a genomic region, the SWI/SNF complex co-activates transcription in part by hydrolyzing ATP via the Snf2 subunit and remodeling nucleosomes to make promoter sequences available to be bound by general transcription factors (TFs) such as TFIID. Some other histone modifying complexes that function in promoting transcription include the NuA4 histone acetyl transferase that acetylates nucleosomal H4 [3], and the Set1 and Set2 histone methyl transferases that methylate nucleosomal H3K4 and H3K36 [4] respectively. Additional protein domains that function in transcription complexes by recognizing some of the diverse histone post-translational modifications (PTMs) include the methyl lysine recognizing PHD and chromodomains [5]. Other ATP dependent chromatin remodeling complexes that function in transcription include the SWR complex that exchanges core H2A in the nucleosome for the transcription-friendly histone H2A variant Htz1 [6,7] and the INO80 complex one function of which is to catalyze the reverse reaction [8].

A typical eukaryotic nucleus is composed of regions of transcriptionally inert heterochromatin as well as euchromatic areas which are considered competent for transcription. The ciliate protozoan *T. thermophila* is a unique model system for studying transcription since it segregates germ-line specific silent (micronucleus-MIC), and somatic transcriptionally active (macronucleus-MAC) chromatin into two distinct nuclei contained within its single cell. The different chromatin structures of the MAC and MIC have their origins in the sexual phase (conjugation) of the life cycle [9]. After pairing, the MIC in each of the two cells undergoes meiosis, generating four haploid meiotic products, only one of which is retained. This gametic nucleus divides mitotically, and one of the two resulting identical haploid nuclei is reciprocally exchanged and fuses with that of its partner to form a genetically identical diploid zygotic nucleus in each cell. The zygotic nucleus divides twice, resulting in four identical products at which point two begin to develop into new MACs (NM). MAC development in the NM of each exconjugant involves extensive programmed DNA rearrangements/irreversible genome silencing that are directly linked to ncRNA-based changes in chromatin structure. These DNA rearrangements include site-specific chromosome fragmentation as well as the deletion of MIC-limited sequences called Internal Eliminated Sequences (IESs), that together result in the loss of ~15% of the germ-line genome [10]. IES deletion begins with the bidirectional transcription of RNAs from the meiotic MIC [11,12]. Meiosis is the only stage of the *Tetrahymena* life cycle where the MIC is transcribed [11,13]. This meiotic MIC-specific transcription is catalyzed by RNAPII [13]. A global MIC-specific nuclear run-on analysis showed that meiotic MIC-specific transcription is biased towards IES DNA, implying that initiation/start-site selection of the MIC-specific transcription is regulated and not simply a result of global or random transcription [12,14]. The underlying molecular mechanisms underlying any transcription in *Tetrahymena* remain poorly understood.

We previously characterized a SNF2-related gene in *T. thermophila* [15]. Despite high primary sequence similarity of Brg1$^{Tt}$ to the budding yeast Snf2 and human Brg1 through most of the protein, Brg1$^{Tt}$ does not possess a recognizable BRD, and its C-terminal region, unlike the entire protein, is dispensable for growth and development [15] raising the possibility that SWI/SNF$^{Tt}$ functions independently of histone acetylation. Here we report a unique BRD-containing protein, Ibd1, which is a component of SWI/SNF$^{Tt}$ during vegetative growth but not during conjugation. Recombinant Ibd1 recognizes several Kac marks on histones that are correlated with active transcription in *Tetrahymena*. AP-MS analysis of Ibd1 revealed it to interact with protein complexes in addition to SWI/SNF$^{Tt}$ including SWR$^{Tt}$, SAGA$^{Tt}$, as well as with a novel putative H3K4-specific histone methyltransferase. ChIP-Seq analysis of Ibd1 suggests a role for the protein during transcription. We suggest that Ibd1 coordinates high levels of transcription of highly expressed genes in *T. thermophila*.

**RESULTS**

**Identification of *T. thermophila* SWI/SNF Complex**

We previously cloned and characterized the Snf2/Brg1 ortholog in *T. thermophila* [15] and predicted it to be a component of a SWI/SNF complex, similar to the situation in *Saccharomyces cerevisiae* [16] and human cells [17]. We used an affinity purification coupled to mass spectrometry (AP-MS) to identify *T. thermophila* SWI/SNF. Specifically, we profiled and compared the set of interacting proteins of two distinct putative SWI/SNF$^{Tt}$ components, Snf5$^{Tt}$ (TTHERM_00304150), a core subunit of yeast and human SWI/SNF complexes [18], and Snf5$^{Tt}$-interacting protein Saf5$^{Tt}$ (TTHERM_00241840). Our comparative sequence analysis shows Snf5$^{Tt}$ to be highly similar to that of yeast and animal cells across most of the protein (see Additional file 1). We generated stable *T. thermophila* cell lines expressing FZZ-epitope tagged *SNF5$^{Tt}$* and *SAF5$^{Tt}$* from their respective macronuclear chromosomal loci by homologous recombination mediated gene replacement [22]. The FZZ epitope tag contains two protein A moieties and one 3xFLAG separated by a TEV cleavage site [23], permitting tandem affinity purification of an FZZ fusion protein, which permits subsequent analysis of co-purifying proteins by Western blotting and/or mass spectrometry [24]. The *SNF5$^{Tt}$-FZZ* and *SAF5$^{Tt}$-FZZ* tagging constructs (see Additional file 1 and 2) were used to transform growing *T. thermophila* strains using biolistic transformation. Gene replacement of the WT *SNF5$^{Tt}$* and *SAF5$^{Tt}$* that occurs by homologous recombination [25] and 'phenotypic assortment' (reviewed in [26]) generates homozygocity in the polyploid MAC for the chromosome containing the *SNF5$^{Tt}$-FZZ* or *SAF5$^{Tt}$-FZZ* gene locus. Western blotting using an FZZ-specific antibody demonstrated expression of the epitope-tagged Snf5$^{Tt}$ or Saf5$^{Tt}$ in whole cell extracts from Snf5$^{Tt}$-FZZ- and Saf5$^{Tt}$-FZZ expressing strains, respectively (Figure 1A, left panel, lanes 2 and 4; and 1B, lanes 3 and 4) compared to that of untagged strains (Figure 1A, left panel, lanes 1 and 3; and 1B, lanes 1 and 2). Indirect immunofluorescence on Snf5[Tt]-FZZ and Saf5[Tt]-FZZ in growing *T. thermophila* showed localization to the transcriptionally active MAC and not to the silent MIC (Figure 1C), identical to what we observed previously for Brg1[Tt] [15], consistent with the hypothesis that Snf5[Tt] and Saf5[Tt] is a member of a Brg1[Tt]-containing SWI/SNF[Tt]. A Brg1[Tt]-specific antibody [15] demonstrated co-purification of Brg1[Tt] with Snf5[Tt]-FZZ and Saf5[Tt]-FZZ affinity purified from whole cell extracts from Snf5[Tt]-FZZ-expressing (Figure 1A, lanes 3-6) and Saf5[Tt]-FZZ-expressing (Figure 1B) but not from untagged strains during vegetative growth.

We next performed a gel-free LC-MS/MS based analysis for each of Snf5-FZZ and Saf5[Tt]-FZZ of the respective affinity purifications to define their sets of interacting proteins. To provide statistical rigor to our AP-MS analyses, all interaction data were filtered using Significance Analysis of INTeractome express (SAINTexpress) which uses semi-quantitative spectral counts to assign a confidence value to individual protein-protein interactions [27]. Application of SAINTexpress to the AP-MS data for two biological replicates of Snf5[Tt]-FZZ and Saf5[Tt]-FZZ affinity purifications from vegetatively growing *T. thermophila* filtered against numerous control AP-MS experiments revealed sets of interaction partners that pass the cut-off confidence value and are listed in Table 1. Our previous analysis [15] of the sequenced *T. thermophila* MAC genome predicted the existence of three potential SWI/SNF proteins in addition to Brg1[Tt] and Snf5[Tt]; Swi1[Tt] (TTHERM_00243900), Swi3[Tt] (TTHERM_00584840), and Snf12[Tt] (TTHERM_00925560). The SAINTexpress analysis of the MS data for Snf5[Tt]-FZZ and Saf5[Tt]-FZZ (Table 1) revealed the identification of the respective baits and each other, in addition to Brg1[Tt], consistent with Figure 1A and 1B, Swi1[Tt], Swi3[Tt], and Snf12[Tt] (Table 1). Saf5[Tt] possess two tandem plant homeodomains (PHD domain). One known function of PHD domains is to mediate specific interactions with methylated lysine on histone proteins to positively regulate transcription [19]. PHD domain-containing proteins are not known to be present in core yeast SWI/SNF but are observed in several animal SWI/SNF complexes [20]. The two PHD domains of Saf5[Tt] are in the same position and are highly similar to those of zebrafish DPF3 and mammalian proteins mBAF45a and hBAF45a (see Additional file 2) both of which are members of a cell type specific SWI/SNF complex [20,21]. DPF3 is part of the BAF chromatin remodeling complex in zebrafish and it is involved in regulation of muscle development and recognizes histones carrying both specific histone acetylation and methylation marks [21]. Snf5[Tt]-FZZ additionally co-purified Tetrin A (TTHERM_00006320), an insoluble cytoskeletal protein unique to ciliates [28]. We have previously noted a variable affinity of the M2 anti-FLAG antibody for this protein as was previously observed for other cytoskeletal proteins [29] and therefore decided not to follow-up on it here. Both Snf5[Tt]-FZZ and Saf5[Tt]-FZZ co-purified with 5 other proteins with no clear orthologs in other described SWI/SNF complexes. The first of these 5 proteins, Saf1 (**S**WI/SNF **A**ssociated **F**actor **1**, Table 1), is predicted to have a coiled coil and a transmembrane domain. Saf1 appears to have a homolog in *Paramecium tetraurelia* (XP_001441480.1) that also possesses the coiled coil domain but not a transmembrane domain. The next 3 proteins, Saf2[Tt], Saf3[Tt], and Saf4[Tt] are *T. thermophila*-specific, meaning that they do not have identifiable known homologs in any other organism. However, all three possess clusters of glutamines in their primary sequence suggestive of a role in transcription [30]. The fifth protein SAINTexpress analysis revealed to co-purify with Snf5$^{Tt}$-FZZ and Saf5$^{Tt}$-FZZ is TTHERM_00729230 (Table 1), which possesses a canonical BRD. We named this protein Ibd1 (**I**nteractive **B**romo**D**omain protein **1**). We suggest the 11 proteins Swi1$^{Tt}$, Swi3$^{Tt}$, Snf5$^{Tt}$, Snf12$^{Tt}$, Brg1$^{Tt}$ and Ibd1 in addition to Saf1-5$^{Tt}$, together define the first known ciliate SWI/SNF complex.

**Ibd1 and BRD-Containing Proteins in *T. thermophila***

The BRD is highly conserved across eukaryotic species, present in functionally diverse proteins including histone acetyl transferases (HATs), ATP-dependent chromatin remodeling complexes, helicases, methyl transferases and transcriptional regulators [31]. Dysfunctional BRD-containing proteins have previously been linked to the development of several human pathologies and are now actively pursued as therapeutic targets [32] . Our finding that a unique BRD-containing protein co-purifies with Snf5$^{Tt}$-FZZ and Saf5$^{Tt}$-FZZ prompted us to determine the full repertoire of BRD-containing proteins in *Tetrahymena.* Our query for BRDs in the *Tetrahymena* genome database, www.ciliate.org [33], identified 14 proteins (Figure 2A). Consistent with human BRD-containing proteins [34], the *Tetrahymena* putative BRD-containing proteins appear functionally diverse and their BRDs can be found in combination with a variety of other domains (Figure 2A). However, unlike humans and yeast, where multiple BRDs can be present within the same protein [34], [35], the *T. thermophila* BRDs are present as single copy. To classify the *T. thermophila* BRD-containing proteome, we carried out a phylogenetic analysis and categorized the set of proteins into three groups based on their BRD similarity (Figure 2B). 'Group I' contains two proteins, Mll1 and BroP-3. The 'Group II' (Figure 2B) can be further categorized into two subgroups such that 'Group II-A' contains only two proteins including Chd1$^{Tt}$ and BroW1$^{Tt}$, whereas 'Group II-B' has six proteins including Snf5$^{Tt}$-interacting Ibd1 (or BroP5; see figure legend for nomenclature). Five out of the 6 proteins found in 'Group II-B' contain no recognizable domains other than BRDs (Figure 2A-B). The similarities in the domain architecture and grouping pattern suggests that the 'group II-B' proteins (which includes Ibd1) might be functionally more similar to each other than to those found within the other groups. Group III contains four proteins including Gcn5$^{Tt}$ and three proteins that possess an ET (extra-terminal) domain in addition to a BRD. In many eukaryotes, including yeast and humans, bromo-domain proteins containing two BRDs followed by an ET domain are referred to as the BET protein family [36]. BRDs generally function to recognize Kac motifs on histones or non-histone proteins to regulate various cellular processes including transcription [34]. The ET domains in contrast are thought to recruit effector proteins which in turn can regulate the transcriptional activity [37]. Structural conservation of a protein often yields insights into its functions. To gain insight into the function of Ibd1, we predicted the three-dimensional structure of its BRD and observed that it folds similarly to the known BRD structures. For example, the predicted structure can be superimposed to the C-terminal BRD of human SMARCA2 (Figure 2C). This suggests that the Ibd1 protein may have a similar function in transcription to that of canonical SNF2 proteins in the yeast and animal SWI/SNF complex through recognition of a similar/same Kac substrate in histones.

**Ibd1 Recognizes Kac and Interacts with Multiple Chromatin Related Proteins**

Our finding of a distinct BRD-containing protein in SWI/SNF$^{Tt}$ is consistent with the fact that a BRD in the catalytic subunit (Snf2/Brg1) has important functions in eukaryotic SWI/SNF complexes. We aligned the primary sequence of the BRD of Ibd1 to those of Gcn5$^{Tt}$, yGcn5p, yBDF1, yBDF2 and ySWI2/SNF2 which are functional BRD-containing proteins (see Additional file 3). The alignment showed a number of conserved amino acids in the BRD including the highly conserved asparagine (N) that makes contact with Kac [34,38] suggesting that Ibd1 as other BRD-containing proteins is likely to bind this mark. We expressed, purified, and incubated recombinant 6xHIS-Ibd1 with a commercially available peptide array that includes a large number of possible histone post-translational modifications, including many histone acetylation sites. Recombinant 6xHIS-Ibd1 displayed strong specificity for acetylated H3K9 and H3K14, acetylated H2AK9 and H2AK13 and tri-acetylated H4K5, H4K8 and H4K12 (Table 2, see Additional file 5 for Raw Data), which are all acetylation patterns associated with the transcriptionally active MAC in *T. thermophila* [39,40]. When incubated on the same peptide array, control recombinant histone methyltransferase 6xHIS-G9a recognized mono- and di-methylated H3K9 (Table 2, see Additional file 5 for Raw Data), as previously demonstrated [41].

We generated a stable line expressing Ibd1-FZZ from its MAC locus. The *IBD1-FZZ* tagging construct (see Additional file 2) was used to transform growing *T. thermophila* strains using biolistic transformation. After selection and phenotypic assortment, Western blotting demonstrated expression of Ibd1-FZZ in whole cell extracts of transformed strains (Figure 3A). Similar to Snf5$^{Tt}$-FZZ, Ibd1-FZZ also co-purifies with Brg1$^{Tt}$ as assessed by Western blotting of affinity purified material (Figure 3B). Gel-free

LC-MS/MS based analysis on affinity purified proteins identified 28 high confidence Ibd1-FZZ co-purifying proteins (Table 3). Comparison of the interaction partners recovered from the purification of Snf5$^{Tt}$-FZZ, Saf5$^{Tt}$-FZZ and Ibd1-FZZ interacting proteins (Figure 3C, Table 3), showed 11 common proteins that co-purify with Ibd1, Saf5$^{Tt}$ and Snf5$^{Tt}$ including Swi1$^{Tt}$, Swi3$^{Tt}$, Snf5$^{Tt}$, Snf12$^{Tt}$ and Brg1$^{Tt}$, Ibd1, and Saf1-5$^{Tt}$ that together we hypothesize form a putative *T. thermophila* SWI/SNF complex. The other 17 high-confidence Ibd1 interacting proteins (Figure 3C, Table 3) could be divided in 3 groups, based on similarity to predicted *S. cerevisiae* orthologs: 1] the SAGA$^{Tt}$ histone acetyl transferase co-activator complex containing Gcn5$^{Tt}$, Ada2$^{Tt}$ and a PhD-containing protein, designated Aap1$^{Tt}$ (**A**da2-**A**ssociated **P**rotein **1**), 2] the SWR$^{Tt}$ ATP-dependent chromatin remodeling complex that in yeast and human cells deposits histone variant Htz1/H2A.Z onto chromatin (Swr1$^{Tt}$, Yaf9$^{Tt}$, Rvb1$^{Tt}$, RvB2$^{Tt}$, Swc2$^{Tt}$ and Swc4$^{Tt}$, Swc5$^{Tt}$ (C-terminal BCNT domain), two actin-like, and three predicted **S**wc4-**A**ssociated **P**roteins (Sap1-3)$^{Tt}$, one of which possess an AT-hook (Sap1$^{Tt}$), the other two (Sap2$^{Tt}$ and Sap3) contain no recognizable domains; and 3] a putative H3K4 methyl transferase (Atrx3/Set1-like). Sap3$^{Tt}$ shares similarity only on a small portion of the protein with hypothetical proteins in *P. tetraurelia* and *Pseudocohnilembus persalinus.* Sap4$^{Tt}$ shares similarity throughout the entire protein with a hypothetical protein in *P. persalinus.* The Ibd1 protein therefore appears to be a component of several chromatin remodeling complexes (SWI/SNF$^{Tt}$, SAGA$^{Tt}$, SWR$^{Tt}$) and one containing an Atrx3/Set1-like HMT.

To further delineate the Ibd1 protein interaction network, we generated separate stable lines expressing Ada2$^{Tt}$-FZZ and Swc4$^{Tt}$-FZZ from their respective MAC loci following an identical strategy as outlined above. SAINTexpress analysis of AP-MS data from growing cells showed that Ada2$^{Tt}$ co-purifies with Ibd1 in addition to the Ibd1-interacting Aap1$^{Tt}$ and Gcn5$^{Tt}$. Additionally, Ada2$^{Tt}$ co-purified with three PHD domain-containing proteins (Aap2$^{Tt}$, Aap3$^{Tt}$ and Aap4$^{Tt}$; Figure 3C, Table 3) and four *T. thermophila*-specific hypothetical proteins (Aap5$^{Tt}$, Aap6$^{Tt}$, Aap7$^{Tt}$ and Aap8$^{Tt}$; Figure 3C, Table 3) that we did not find to co-purify with Ibd1-FZZ. We suggest that the Ada2 interacting proteins together represent a *Tetrahymena* SAGA$^{Tt}$ complex (Figure 3C, Table 3). SAINTexpress analysis of Swc4$^{Tt}$-FZZ AP-MS revealed it to co-purify a subset of Ibd1 interacting proteins that were predicted to be SWR$^{Tt}$ complex proteins (Figure 3C, Table 3). Swc4$^{Tt}$-FZZ further interacts with *T. thermophila* orthologs of the Tra1 and Tra2 PI3 kinases (Figure 3C, Table 3), neither of which co-purified with Ibd1. In yeast, Swc4 co-purifies with Tra1 via the NuA4 histone acetyltransferase complex of which Swc4 is a component, in addition to SWR-C. We did not observe Swc4$^{Tt}$-FZZ to co-purify with any protein that would indicate it to be a member of a *T. thermophila* NuA4 complex. The set of proteins that we hypothesize to constitute SWR$^{Tt}$ are listed in Table 3. Although the *T. thermophila* genome encodes a predicted ortholog of Swc6/Vps71 (TTHERM_01298590) we did not find it to co-purify with Swc4$^{Tt}$ or Ibd1 in growing cells.

**Ibd1 Function During Conjugation**

To gain further insight into Ibd1 function, we assessed its expression through growth and sexual development. We performed Western blotting of whole cell extracts made at different times during the *T. thermophila* life cycle, probing for Ibd1-FZZ (Figure 4A, lower panel). We have previously demonstrated Brg1$^{Tt}$ to have relatively constant levels of expression throughout growth and development [15]. We therefore used anti-Brg1$^{Tt}$ as a loading control (Figure 4A, top panel), and anti-Pdd1 [42] as a development-specific control (Figure 4A, middle panel) for these experiments. Similar to Brg1$^{Tt}$, Ibd1 is expressed throughout the *T. thermophila* life cycle. Indirect immunofluorescence of Ibd1-FZZ performed on growing and conjugating cells (Figure 4B) demonstrated localization exclusively to the MAC during growth and conjugation, specifically to the parental MAC through early nuclear development including meiosis (Figure 4B: 0-6h) before switching to the anlagen mid-way through sexual development (Figure 4B:8h). This is similar to what was shown previously for Brg1$^{Tt}$ [15]. In particular, as for Brg1$^{Tt}$, localization of Ibd1-FZZ in the parental macronucleus is lost at the onset of macronuclear development, a stage where the two anterior nuclei (the anlagen) have become visibly larger than the posterior nuclei (Figure 4B Ibd1-FZZ cells, compare 8h and 6h post-mixing). The cellular localization of Ibd1 is therefore correlated with transcriptionally active MAC during growth and nuclear development.

To determine if Ibd1's protein interaction network changes during sexual development, we performed AP-MS using whole cell extracts prepared from conjugating cells harvested 5 hours post-mixing, a time period following meiosis that is marked by a series of rapid post-zygotic nuclear divisions and where Ibd1-FZZ is found exclusively in the parental MAC (Figure 4B). SAINT-curated AP-MS data are shown in Additional File 6. Direct comparison of the Ibd1 AP-MS results from vegetative and conjugating cells revealed that members of the SWI/SNF$^{Tt}$ and the Atrx3$^{Tt}$/Set1$^{Tt}$ HMT complexes were associated with Ibd1-FZZ to a lower degree in conjugation than during vegetative growth, while members of the putative SWR$^{Tt}$ and SAGA$^{Tt}$ remained relatively unaffected (Figure 5A). The recovery as defined by spectral counts of SWI/SNF$^{Tt}$ members (Figure 3C, Table 1, Table 3) appeared relatively low at this stage when compared to members of SWR$^{Tt}$ and SAGA$^{Tt}$. To validate this finding, we used M2 agarose to affinity purify Ibd1-FZZ from untagged and Ibd1-FZZ expressing cells and blotted with anti-Brg1 antibody following SDS-PAGE (Figure 5B). In these conjugating cells, Ibd1-FZZ did not co-purify with Brg1$^{Tt}$ (Figure 5B), consistent with the substantially lower amounts of the protein detected by mass spectrometry. These data suggest a profound modulation of the Ibd1 interactome favoring its association with SWR$^{Tt}$ and SAGA$^{Tt}$ over SWI/SNF$^{Tt}$ complex early in conjugation (5 hours post-mixing).

**Ibd1 Localizes to Transcriptionally Active Chromatin**

As noted above, Ibd1 co-purifies with multiple protein complexes involved in gene expression regulation and *in vitro* recognizes histone marks associated with an active chromatin state. These observations suggest an intimate role of Ibd1 in transcription regulation. To examine this possibility in more detail, we employed chromatin immunoprecipitation followed by next generation sequencing (ChIP-Seq). Specifically, we asked whether Ibd1 localizes to specific regions of the genome that correlate with transcriptionally active chromatin.

Data for two biological replicates that include DNA from input chromatin as well as Ibd1-FZZ precipitate from two independent experiments were analyzed. Our ChIP-Seq (GEO accession GSE103318) data set utilizing the available genome annotations [33] was composed of all annotated genic or open reading frames (ORF) and intergenic regions. The two generated lists displayed greater than or equal to 2-fold enrichment of Ibd1 and were ranked in descending order (see Additional Files 7 and 8, All_>2X_Fold_Enrichment tab). From these lists we observed that Ibd1 strongly occupies to 837 ORF and 396 intergenic regions with an enrichment (IP/INPUT) greater than or equal to 2-fold (Figure 6A, see Additional Files 7 and 8, >2X_Enriched_with_Strong_Peaks tab). We initially focused our attention to the identified 837 ORFs and assessed the transcriptional state of these genes. We utilized previously published RNA-Seq data that has been used to rank genes based on their expression level during vegetative growth (GEO accession GSM692081,[43]). Based on this data we found that 9% and 29% of genes in *Tetrahymena* are highly and moderately expressed respectively (Figure 6B, left panel, see Additional file 7, RNA-Seq tab). On the other hand, we found that 54% (457 ORF) and 16% (134 ORF) of genes occupied by Ibd1 are highly and moderately expressed respectively (Figure 6B, right panel, and Figure 6C, see Additional file 7, localization tab). These observations are consistent with our histone peptide- array data and further strengthen the idea that Ibd1primarily occupies active chromatin regions. Interestingly, Ibd1 showed binding to 114 ORF with low expression to no-expression during vegetative growth. (Figure 6C, see Additional file 7, localization tab). The overall trend of the Ibd1 binding pattern to highly expressed genes that are highly occupied is particularly evident for genes that have enrichment greater than or equal to 4-fold (298 genes in total) (Figure 6C). To examine whether these 298 genes are enriched for any particular functional categories, we grouped them using STRING [44] based on their predicted Gene Ontology (GO) terms [45]. We identified 122 genes that are significantly enriched with a particular term related to housekeeping functions, such as biological process, cellular process, translation, metabolic processes and gene expression (Figure 6D, see Additional file 7, 4X+_GO_Biological_Expression tab). These housekeeping genes are generally highly expressed consistent with our findings that Ibd1 primarily occupies transcriptionally active chromatin. To compare this data with the overall distribution of all *Tetrahymena*'s annotated genes the same approach was used (Figure 6E, see Additional file 7, AllTtGenes_GO_Biological_Proces tab). Figure 6D and 6E suggest that Ibd1 mainly controls housekeeping genes in vegetative cells.

To validate our ChIP-Seq analysis of Ibd1 enriched chromatin, we designed primers for the three genes that showed the highest Ibd1-FZZ fold enrichment (see Additional file 7,>2X_Enriched_with_Strong_Peaks tab) as well as a fourth, *PDD1* which is exclusively developmentally expressed [46] and did not show enrichment for Ibd1-FZZ during growth (see GEO accession GSE103318). Our ChIP-qPCR analysis of the four genes confirmed specific enrichment of Ibd1-FZZ in *HTA3*, *RPS22*, and *HFF1* but not *PDD1* relative to chromatin made from untagged cells (Figure 6F, see Additional file 9 for Raw data). We conclude that Ibd1 occupies transcriptionally active chromatin and might have a role in regulating the expression of a subset of genes involved in basal cellular housekeeping functions.

**Localization of Ibd1 in *Tetrahymena*'s genome**

We next examined our ChIP-Seq data for both ORFs and intergenic regions that showed greater than or equal to 4-fold enrichment to determine how Ibd1 is situated in the genome relative to ORF and intergenic regions.

Using this fold-enrichment cut-off, we obtained 298 genic and 140 intergenic regions.

We first investigated the genic regions to assess the Ibd1 peak distribution. Figure 7A shows a representative example of Ibd1 ORF-specific localization where peaks are primarily enriched within the gene-body (see Additional file 7, 4X_+_Ibd1_Occupancy tab for the full list). Next, to classify 140 intergenic regions, we manually inspected the ChIP-Seq peaks using the genome browser [47] and categorized them into five groups based on their localization (Fig 7B-F, see Additional file 8, Intergenic_Groups tab). The promoter group showed intergenic localization that was proximal to the 5' region of 91 single predicted genes (e.g. Figure 7B). The Ibd1 terminator group showed intergenic localization proximal to the 3' region of 33 single predicted genes (e.g. Figure 7C). The third intergenic group showed Ibd1 localization to 2 regions where there is an overlap between the promoter of one predicted gene and the terminator of another (e.g. Figure 7D). The fourth group showed localization of Ibd1 to 13 single 5' promoter regions potentially controlling expression of two predicted genes (Figure 7E). The fifth group showed localization of Ibd1 to 11 single terminator 3' regions of two distinct predicted genes (Figure 7F). We found that among the 298 ORF showing ≥4X Ibd1 enrichment, 37 also additionally showed enrichment through the promoter (Figure 7G and Additional file 8, Combining_Intergenic_and_ORF tab for list) and 19 at the terminator region (Figure 7H and Additional file 8, Combining_Intergenic_and_ORF tab). Collectively these data suggest that Ibd1 appears to bind near the promoters and within gene bodies, consistent with a role in transcription regulation through its potential role in organizing multiple protein complexes.

**DISCUSSION**

**Ibd1 is a BRD-containing protein that interacts with multiple chromatin remodeling complexes in *T. thermophila***

In our previous molecular characterization of Brg1$^{Tt}$ [15], we reported that it lacked a C-terminal BRD which differs from the case in yeast (Snf2/Sth1) and mammalian cells (Brg1/Brahma). We report here that a distinct, BRD-containing protein, Ibd1, is a member of the *Tetrahymena* SWI/SNF complex. Recombinant Ibd1 recognized several Kac histone PTMs that are correlated with transcription. Ibd1 however established a large interaction network beyond the SWI/SNF$^{Tt}$ complex including putative SAGA$^{Tt}$ and SWR$^{Tt}$ complexes as well a Atrx3/Set1-like HMT that is predicted to be H3K4 specific, a modification linked to transcription. As is standard practice, we used a promiscuous DNAse and RNAse (benzonase nuclease) in the preparation of whole cell extracts used for AP-MS (as detailed in Materials and Methods). Very little, if any nucleic acid remains in our extract submitted to AP-MS. Also, although Ibd1 AP-MS yielded several putative protein complexes, reciprocal purification of individual complex components co-purified Ibd1 but not the other complexes consistent with binding of other proteins to Ibd1 being specific and independent of DNA. This being said, we cannot exclude that nucleic acids already bound by proteins are protected from nuclease cleavage and may contribute to the observed binding events,

**Characterization of a *Tetrahymena* SWI/SNF complex**

The *Tetrahymena* SWI/SNF$^{Tt}$ complex, as defined by the set of proteins that co-purify with Ibd1$^{Tt}$, Snf5$^{Tt}$, and Saf5$^{Tt}$, includes orthologs of canonical SWI/SNF proteins Swi1,

Swi3, Snf5, Snf12, and Snf2/Brg1, the PHD domain-containing Saf5, as well as several ciliate and species- specific novel proteins. Of note, three of the novel proteins that co-purify with *Tetrahymena* SWI/SNF (Saf2[Tt], Saf3[Tt] and Saf4[Tt]) do not possess conserved domains outside of glutamine-rich regions. Yeast and mammalian Swi1[Tt] possess an AT-rich Interactive (ARID) and also a Q-rich domain [48]. Swi1[Tt] possesses an ARID but not a Q-rich domain. We suggest that in SWI/SNF[Tt], the Q-rich proteins Saf2[Tt], Saf3[Tt], and Saf4[Tt] act in conjunction with Swi1[Tt]. The Q-rich domain in animal Sp1 functions as an activation domain for transcription factors through recruitment of general transcription factor(s) [49]. We suggest that the function of Saf2-4[Tt] is to function in co-activation by recruiting general transcription factors and/or RNA polymerase to promoter regions of highly expressed genes in growing *Tetrahymena*.

The finding that Ibd1 is a member of SWI/SNF[Tt] is informative in that its BRD interacts with Kac of histone proteins, similar to that observed for Snf2/Sth1 in yeast [50] and Brahma/Brg1 in humans [51]. In addition to the BRD-containing Ibd1, *Tetrahymena* SWI/SNF also contains a PHD domain-containing protein, Saf5[Tt]. One function attributed to PHD domains is recognizing methylated lysines in proteins such as histones. For example the PHD domain of human ING2 recognizes H3K4me3 [52]. Thus, the SWI/SNF[Tt] contains two proteins that potentially recognize PTM on histones, Saf5[Tt] that likely recognizes methyl lysine (and possibly acetyl-lysine [21]), and Ibd1 that recognizes Kac. The *Tetrahymena* transcriptionally active MAC contains hyper-acetylated histone H3 that is also di- or tri-methylated on H3K4 [40]. We suggest that a subset of these modified H3-containing nucleosomes can be recognized by SWI/SNF which would then remodel them to facilitate transcription. Additional SWI/SNF co-activator function could be derived from recruitment of general TFs and/or RNA polymerase II by the Saf2-4 proteins with Q-rich regions. Ibd1 may not interact with SWI/SNF in development in the same manner as it does during vegetative growth. We suggest that the function of SWI/SNF during nuclear development occurs independent of histone acetylation.

**Tetrahymena Ibd1-containing SWR, SAGA and HMT complexes**

In addition to being a member of SWI/SNF$^{Tt}$, Ibd1 is also a distinct component of the SWR and SAGA complexes as well as interacting with an uncharacterised H3K4-specific histone methyl transferase that is similar to human Atrx3 and yeast Set1. The function of the SWR complex in fission [53] and budding [6] yeasts is the deposition of the histone H2A variant Pht1/Htz1 (H2A.Z in humans, and Hv1 in *Tetrahymena*). Deposition of Htz1 in budding yeast is linked to NuA4-dependent histone acetylation via the BRD-containing Bdf1 subunit of SWR [54]. In yeast, Bdf1 is also a component of TFIID linking histone acetylation to pre-initiation complex assembly [55]. In *Tetrahymena*, Ibd1 did not co-purify with any proteins similar to components of the general transcription apparatus. Like Ibd1, Hv1 is localized to transcriptionally active MAC in growing cells [56]. Unlike Ibd1, Hv1 localizes also to the crescent MIC corresponding to meiotic prophase [57], a time period in *Tetrahymena* where large genome-wide transcription of the MIC by RNAPII occurs (reviewed in [58]).

In budding yeast, SWR is functionally linked to the NuA4 histone acetyl transferase complex via shared subunits Swc4 and Yaf9. In *Tetrahymena,* Swc4$^{Tt}$ did not co-purify with a histone acetyl transferase subunit and may not be a member of a NuA4-type complex. In fact, a strict NuA4 type complex in *Tetrahymena* is unlikely to exist, despite the presence of 3 genes encoding MYST family histone acetyl transferases. A previous study did identify a H2A/H4 nucleosomal HAT similar to the activity of NuA4 but also showed by glycerol gradient analysis that the activity purifies at ~80kD [59]. Consistent with this observation, the MAC does not appear to encode a gene that is a clear ortholog of the conserved NuA4 subunit such as Epl1/EPC so it is unclear if there exists a "piccolo" NuA4 [60]. Swc4$^{Tt}$ did co-purify with orthologs of Tra1$^{Tt}$ and Tra2$^{Tt}$ kinases that did not purify with Ibd1 (Table 3 and Figure 3C). In *S. cerevisiae* Tra1 co-purifies with NuA4 [61] and SAGA [62] that contribute to their co-activator function [63]. It will be interesting to determine whether SAGA$^{Tt}$ fulfils the function of SAGA and NuA4 in budding yeast or whether there exists a divergent version of NuA4 in *Tetrahymena*. Ibd1 co-purifies with Gcn5$^{tt}$ and Ada2$^{Tt}$ in addition to the PHD domain-containing A2A1$^{Tt}$. Ada2$^{Tt}$ co-purifies with these proteins in addition to seven others including three additional PHD domain-containing proteins A2A2-4$^{Tt}$. Thus, Ada2$^{Tt}$ co-purifies with four distinct PHD domain-containing proteins. Further work will be necessary to determine whether the set of Ada2 interacting proteins represent a single assemblage or if Ibd1, Ada2 and Gcn5 represent a 'core' to the *Tetrahymena* SAGA complex that can have different specificity depending on which PHD protein it is interacting with at a particular time.

**Model for Ibd1 function**

We hypothesize that Ibd1 has a common function that it performs in diverse chromatin remodeling complexes. Consistent with a function in promoting transcription, Ibd1-FZZ specifically localized to the coding regions of multiple highly transcribed genes during vegetative growth. A model for Ibd1 function is that it recognizes one or more specific histone Kac marks that are associated with transcription and recruits multiple chromatin related complexes to the region to either further acetylate nearby chromatin (SAGA$^{Tt}$), to remodel nucleosomes (SWI/SNF$^{Tt}$), to deposit Hv1 (SWR$^{Tt}$), and to di- or tri-methylate histone H3K4 (Atrx3/Set1-like histone methyl transferase). SWI/SNF, SAGA and SWR, and H3K4 methylation are all linked to transcription in other experimental systems. We predict that Ibd1 is particularly important to maintain high rates of transcription on highly expressed genes such as those encoding the core histones or ribosomal proteins. Our ChIP-Seq analysis of Ibd1 supports this hypothesis with strong occupancy of the coding regions of genes encoding core histones HHT1 and HHF1. ChIP-Seq of Ibd1-containing complex specific members (i.e. Snf5$^{Tt}$, Swr1$^{Tt}$, Ada2$^{Tt}$, etc.) will be required to test the validity of this hypothesis. As well as being found in coding regions, Ibd1 also localizes to the regulatory region of several genes. Further work will be necessary to determine whether Ibd1 is necessary for the recruitment of SWI/SNF$^{Tt}$, SAGA$^{Tt}$, SWR$^{Tt}$, and the HMT to ORFs and the regulatory regions identified in our ChIP seq analysis. It will also be interesting to determine whether the regulatory regions enriched in Ibd1 contain conserved DNA sequences that may indicate whether specific DNA-binding transcription factors recruit Ibd1-containing protein complexes to regulatory regions.

**BRD proteins in *Tetrahymena***

We have identified and performed a phylogenetic analysis on 14 BRD-containing proteins in *Tetrahymena*. Ibd1 is a member of a grouping that includes six proteins, five of which are like Ibd1 in possessing a single BRD and no other recognizable domains. Four of these 5 are similar in length to Ibd1 suggesting relatively recent evolutionary divergence of the four. BRD inhibitors are currently of significant clinical interest in the development of drugs to treat parasitic infections as a number of apicomplexan protozoan parasites possess lineage-specific BRD proteins that appear to be important for various stages of their life cycle [64]. Because the ciliates and apicomplexans are closely related in evolution, we suggest *Tetrahymena* may provide a tractable model for molecular analysis of some of these BRD proteins.

**CONCLUSIONS**

In multi-cellular eukaryotes, the precise function of how chromatin remodeling complexes work is poorly understood. Alteration or loss of factors involved in these complexes through mutation has been shown to be associated with cancer. We utilized the protist model, the Aleveloate *Tetrahymena thermophila* which segregates transcriptionally active, and silent chromatin into two distinct nuclei, the macronucleus (MAC) and micronucleus (MIC) respectively, contained in the same cell. Through the discovery of a bromodomain-containing protein, Ibd1, we advanced the knowledge of chromatin remodeling complexes in protists by defining for the first time the protein complements of SWI/SNF, SWR and SAGA complexes. In addition, we present a model where a single protein, Ibd1 coordinates the action of multiple chromatin remodelling complexes to achieve high levels of transcription. Our research will contribute to our current understanding of transcription in ciliates, and more broadly the function and diversity of chromatin remodeling complexes in eukaryotes.

## MATERIALS AND METHODS

### Protein Sequence alignments

Multiple sequence alignments of Snf5, Saf5 and Ibd1 amino acid sequence from various model organisms were performed using ClustalOmega (http://www.ebi.ac.uk/Tools/msa/clustalo/) and then shaded by importing the ALN file into the Boxshade server (http://www.ch.embnet.org/software/BOX_form.html). SMART [65] was used to find the beginning and end of the domains .

### Cell strains

*T. thermophila* strains CU428 [Mpr/Mpr (VII, mp-s)] and B2086 [Mpr+/Mpr+ (II, mp-s)] of inbreeding line B were obtained from the *Tetrahymena* Stock Center, Cornell University, Ithaca N.Y. (http://tetrahymena.vet.cornell.edu/). Cells were cultured axenically in 1× SPP at 30 °C as previously described [66].

### Oligonucleotides

See Additional file 4 for a list of the oligonucleotides used during this study.

### DNA manipulations

Whole-cell DNA was isolated from *T. thermophila* strains as described [67]. Molecular biology techniques were carried out using standard protocols or by following a supplier's instructions.

**Affinity Purification, Sample Preparation and Mass Spectrometric Analysis:**

AP-MS analysis were performed as per [24] with minor modifications, see Additional file 10.

**Macronuclear gene replacement**

Epitope tagging vectors for Snf5, Saf5, Ibd1, Swc4 and Ada2 were constructed as previously described [24].

**ChIP**

ChIP was performed as described [68] with modifications described in Additional file 10.

**NGS**

Sequencing and analysis of DNA co-purifying with ChIP of Ibd1-FZZ is described in Additional file 10.

**ChIP-qPCR**

Four ChIP biological repetitions for the Ibd1-FZZ and three ChIP repetitions for the untagged cell lines were quantified (Nanodrop, Thermo Scientific) and diluted to reach the smallest DNA concentration found in a sample (1 to 3.1ng/μL of DNA). Master mixes with a final volume of 20µL were prepared (Sybr Green Supermix, Cat. #1708880, Bio-Rad) to amplify: the top 3 genes that presented the highest fold enrichment from Ibd1-FZZ ChIP-Seq and are highly expressed and a gene that is not expressing during vegetative growth (*PDD1*) (Primers, see Additional file 4) using qPCR (CFX 96-well Real-Time System, Bio-Rad) with the following parameters: Initial denaturation at 98°C for 3 minutes; 40 cycles of amplification at 95°C for 15 seconds and 60°C for 60 seconds followed by acquisition in the SYBR/FAM channel; and melting curve from 65°C to 95°C increasing 0.5°C/cycle and acquisition every 0.5 seconds in the SYBR/FAM channel. Each targeted gene was considered as an individual experiment each with its own standard curve. The standard curve for each target has 3 points representing 100%, 10%, and 1% of the corresponding input sample. The largest point of the curve was undiluted input sample and was followed by serial dilutions. (see Additional file 9). Raw Cq values for input DNA and IP DNA were analyzed using the BioRad Prime PCR program, which normalizes these data to the generated standard curve that we represented as % with respect to the INPUT. Ultimately, these normalized ChIP data are expressed as fold enrichment, by dividing normalized IP over normalized Input. The standard error of the mean (SEM) was calculated for each duplicate (see Additional file 9).

**Additional experimental procedures can be accessed in Additional file 10**

**DECLARATIONS**

**Availability of data and materials**

Mass spectrometry data have been deposited in the Mass spectrometry Interactive Virtual Environment (MassIVE, http://massive.ucsd.edu) MSV000081461. All MS files used in this study were deposited at MassIVE (http://massive.ucsd.edu) and were assigned the MassIVE identifier MSV000081461. Direct link to MassIVE dataset: http://massive.ucsd.edu/ProteoSAFe/dataset.jsp?task=75098964a529429c943dac8a9ae537f7

ChIP seq data generated in this paper can be found online at Gene Expression Omnibus (GEO, http://www.ncbi.nlm.nih.gov/geo/) GSE103318. NGS and peaks files (Private data, not to be shared or distributed without permission) produced in this study were deposited at https://www.ncbi.nlm.nih.gov/geo/ with unique identifier GSE103318. Direct link: http://www.ncbi.nlm.nih.gov/geo/query/acc.cgi?acc=GSE103318

**Funding**

Work in the Fillingham and Lambert lab was supported by Natural Sciences and Engineering Research Council of Canada (NSERC) Discovery Grant 131034 and 1304616, respectively.  Work in the Pearlman lab was supported by Canadian Institutes of Health Research (CIHR) MOP13347 and Natural Sciences and Engineering Research Council of Canada NSERC) Discovery grant 539509.  Work in the Gingras lab was supported by the Canadian Institutes of Health Research (CIHR) Foundation Grant (FDN 143301). Proteomics work was performed at the Network Biology Collaborative Centre at the Lunenfeld-Tanenbaum Research Institute, a facility supported by Canada Foundation for Innovation funding, by the Ontarian Government and by Genome Canada and Ontario Genomics.  A.-C.G. is the Canada Research Chair (Tier 1) in Functional Proteomics. J.-P.L. was funded by a Scholarship for the Next Generation of Scientists from the Cancer Research Society. SciNet is funded by: The Canada Foundation for Innovation under the auspices of Compute Canada; the Government of Ontario; Ontario Research Fund - Research Excellence; and the University of Toronto.

**Conflict of interest**

The authors declare no conflict of interest.

**Authors' contributions**

AS generated Ibd1-FZZ, Saf5-FZZ and 6xHIS-Ibd1 cell lines, and performed Immunoprecipitations, affinity purifications, peptide array, IF microscopy for Ibd1-FZZ and Saf5-FZZ, ChIP-Seq, ChIP-qPCR, Western Blots, prepared figures and wrote manuscript.  JG generated Snf5-FZZ, Swc4-FZZ, Ada2-FZZ cell lines, and performed IF microscopy for Snf5-FZZ.  JPL processed and analyzed samples for mass spectrometry, generated figures, participated in writing the manuscript and editing.  SNS performed bioinformatics analyses of bromodomains in *Tetrahymena thermophila*.  MP participated in processing and analysis of ChIP-Seq data.  AB participated in peptide array and IF, and CTM participated in ChIP-Seq.  ACG and RP were responsible for supervision, and manuscript editing.  JF conceived the study, participated in its design and coordinated and edited the manuscript. All authors read and approved the final manuscript.

**Ethics approval and consent to participate**

Not applicable

**Consent for publication**

Not applicable

**Competing interests**

The authors declare that they have no competing interests

**Acknowledgements**

We acknowledge Saba Safar and Camila Imamura for technical expertise.

**TABLES**

**Table 1: AP-MS data for Snf5-FZZ and Saf5-FZZ uncovers predicted and novel members of a *Tetrahymena* SWI/SNF complex.** Curated SAINTexpress data from 2 biological replicates of SNF5-FZZ and SAF5-FZZ AP-MS samples. Genes in *italics* were previously predicted to be SWI/SNF components (Fillingham et al. (2006)). Saf (SWI/SNF Associated Factor), Ibd (Interactive Bromodomain Protein). The members of the SWI/SNF complex are the first 11 rows.

| TTHERM | Gene Name | Spectral Count Sum Snf5 (BAIT) | Spectral Count Sum Saf5 (BAIT) | SWI/SNF Yeast Ortholog | SWI/SNF Human Ortholog | Notes |
|---|---|---|---|---|---|---|
| *TTHERM_00584840* | *SWI3* | 308 | 835 | Swi3 | BAF170/SMARCC2 | --- |
| *TTHERM_01245640* | *BRG1* | 264 | 384 | Snf2 | BRM/SMARCA2 | SNF2 catalytic subunit |
| *TTHERM_00243900* | *SWI1* | 140 | 171 | Swi1 | BAF250A/ARID1A | --- |
| *TTHERM_00304150* | *SNF5* | 133 | 72 | Snf5 | BAF47/SMARCB1 | --- |
| *TTHERM_00925560* | SNF12 | 94 | 137 | Snf12 | SMARCD2 | --- |
| TTHERM_00092790 | *SAF1* | 85 | 78 | --- | --- | Transmembrane protein, putative |
| TTHERM_00346460 | *SAF2* | 136 | 97 | --- | --- | Hypothetical protein - 13% Glutamine |
| TTHERM_00129650 | *SAF3* | 79 | 40 | --- | --- | Hypothetical protein - 26% Glutamine |
| TTHERM_00637690 | *SAF4* | 32 | 84 | --- | --- | Hypothetical protein - 31% Glutamine |
| TTHERM_00241840 | *SAF5* | 64 | 56 | --- | BAF45a | PHD finger containing protein |
| TTHERM_00729230 | *IBD1* | 48 | 107 | --- | --- | Bromodomain-containing protein |
| TTHERM_00006320 | Tetrin A | 33 | --- | --- | --- | --- |

**Table 2. Histone peptide array data reveals the top post-translational modification recognized by Ibd1.** The histone peptide array contains human histone modifications that resemble *Tetrahymena*'s histones. The intensity average columns show the top 10 histone modifications recognized by 6xHIS-Ibd1 and 6xHIS-G9a in bold. Shaded means that the amino acid is not present in the *Tetrahymena*'s histone (see Additional file 5 for Raw Data).

| Histone | Modification 1 | Modification 2 | Modification 3 | Modification 4 | Intensity Average 6xHIS-Ibd1 (4 repetitions) | Intensity Average 6xHIS-G9a (2 repetitions) |
|---|---|---|---|---|---|---|
| H3 | **K9ac** | **K14ac** | | | **0.95** | 0.02 |
| H2a | K5ac | **K9ac** | **K13ac** | | **0.92** | 0.01 |
| H4 | R3me2s | **K5ac** | **K8ac** | **K12ac** | **0.90** | 0.00 |

| H4 | **K5ac** | **K8ac** | **K12ac** | K16ac | **0.89** | 0.03 |
| H3 | K9me3 | **K14ac** | | | **0.88** | 0.01 |
| H3 | S10P | **K14ac** | | | **0.88** | 0.04 |
| H4 | **K5ac** | **K8ac** | **K12ac** | | **0.84** | 0.01 |
| H3 | T11P | **K14ac** | | | **0.82** | 0.01 |
| H2a | S1P | K5ac | **K9ac** | **K13ac** | **0.82** | 0.01 |
| H2a | S1P | **K9ac** | **K13ac** | | **0.81** | 0.01 |
| H3 | R2me2s | K4me2 | R8me2a | **K9me2** | 0.19 | **0.96** |
| H3 | R2me2a | K4me1 | R8me2a | **K9me2** | 0.18 | **0.94** |
| H3 | R2me2a | K4me2 | R8me2a | **K9me2** | 0.18 | **0.90** |
| H3 | K4ac | R8me2s | **K9me1** | | 0.21 | **0.84** |
| H3 | R2me2a | K4me2 | R8me2a | **K9me1** | 0.19 | **0.83** |
| H3 | R2me2a | K4me3 | R8me2a | **K9me2** | 0.18 | **0.80** |
| H3 | R2me2a | K4ac | R8me2a | **K9me2** | 0.28 | **0.79** |
| H3 | R2me2a | K4me3 | R8me2a | **K9me1** | 0.19 | **0.79** |
| H3 | R2me2s | K4ac | R8me2a | **K9me1** | 0.16 | **0.76** |
| H3 | R2me2s | K4me2 | R8me2s | **K9me1** | 0.17 | **0.76** |

**Table 3: AP-MS data for Ibd1-FZZ, Ada2-FZZ and Swc4-FZZ purified from vegetative cells.** Curated SAINTexpress data from 2 biological replicates of Ibd1-FZZ, Ada2-FZZ and Swc4-FZZ AP-MS samples.

| TTHERM | Gene Name | Spectral Count Sum Ibd1 (BAIT) | Spectral Count Sum Swc4 (BAIT) | Spectral Count Sum Ada2 (BAIT) | Possible complex | Notes |
|---|---|---|---|---|---|---|
| TTHERM_00729230 | *IBD*1 | 827 | 99 | 58 | All listed below | Bromodomain-containing protein |
| TTHERM_00486440 | Atxr3/*SET1*-like | 125 | --- | --- | COMPASS | --- |
| TTHERM_00584840 | *SWI*3 | 411 | --- | --- | SWI/SNF | --- |
| TTHERM_01245640 | *BRG*1 | 202 | --- | --- | SWI/SNF | SNF2 catalytic subunit |
| TTHERM_00925560 | *SNF*12 | 81 | --- | --- | SWI/SNF | --- |
| TTHERM_00243900 | *SWI*1 | 92 | --- | --- | SWI/SNF | --- |
| TTHERM_00304150 | *SNF*5 | 64 | --- | --- | SWI/SNF | --- |
| TTHERM_00092790 | *SAF*1 | 47 | --- | --- | SWI/SNF | Transmembrane protein, putative |
| TTHERM_00346460 | *SAF*2 | 39 | --- | --- | SWI/SNF | Hypothetical protein – 13% Glutamine |
| TTHERM_00637690 | *SAF*4 | 36 | --- | --- | SWI/SNF | Hypothetical protein – 31% Glutamine |
| TTHERM_00129650 | *SAF*3 | 24 | --- | --- | SWI/SNF | Hypothetical protein – 26% Glutamine |
| TTHERM_00241840 | *SAF*5 | 29 | --- | --- | SWI/SNF | PHD finger containing protein |
| TTHERM_00317000 | Actin-like | 63 | --- | --- | Undefined | --- |
| TTHERM_00046920 | *RVB*2 | 174 | 893 | --- | SWR | --- |
| TTHERM_00476820 | *RVB*1 | 121 | 333 | --- | SWR | --- |
| TTHERM_01546860 | *SWR*1 | 113 | 561 | --- | SWR | --- |
| TTHERM_00975380 | Actin-like | 101 | 542 | --- | SWR | --- |
| TTHERM_01005190 | *ARP*6 | 18 | 82 | --- | SWR | --- |
| TTHERM_00170260 | Sap3 | 23 | 144 | --- | SWR | Hypothetical protein |
| TTHERM_00357110 | *SWC*4 | 33 | 419 | --- | SWR | --- |
| TTHERM_00136450 | *SWC5/AOR*1 | 32 | 121 | --- | SWR | Bucentaur or craniofacial development containing protein |
| TTHERM_00355040 | Sap1 | 20 | 78 | --- | SWR | AT-hook containing protein |
| TTHERM_00388500 | *SWC*2-like | 19 | 75 | --- | SWR | --- |
| TTHERM_00561450 | *YAF*9 | 15 | 200 | --- | SWR | --- |
| TTHERM_00046150 | Sap2 | 9 | 32 | --- | SWR | Hypothetical protein |
| TTHERM_00978770 | *TRA*1 | --- | 767 | --- | Undefined | --- |
| TTHERM_00979770 | *TRA*2 | --- | 179 | --- | Undefined | --- |
| TTHERM_00444700 | Aap1 | 51 | --- | 356 | SAGA | PHD finger containing protein |
| TTHERM_00248390 | *GCN*5 | 32 | --- | 605 | SAGA | --- |
| TTHERM_00790730 | *ADA*2 | 28 | --- | 429 | SAGA | --- |
| TTHERM_00145290 | Aap6 | --- | --- | 152 | SAGA | --- |

| TTHERM | | | | | | |
|---|---|---|---|---|---|---|
| TTHERM_00313140 | Aap2 | --- | --- | 69 | SAGA | PHD finger containing protein |
| TTHERM_00670640 | Aap4 | --- | --- | 314 | SAGA | PHD finger containing protein |
| TTHERM_00502120 | Aap3 | --- | --- | 184 | SAGA | PHD finger containing protein |
| TTHERM_00420400 | Aap8 | --- | --- | 194 | SAGA | --- |
| TTHERM_00046390 | Aap7 | --- | --- | 93 | SAGA | --- |
| TTHERM_00464970 | Aap5 | --- | --- | 247 | SAGA | --- |

**Table 4. Top Ibd1 ChIP-Seq hits during vegetative growth.** The 3-top highly expressed genes and an exclusive developmental gene are shown.

| TTHERM | Description | Fold Enrichment | Highly expressed |
|---|---|---|---|
| TTHERM_00143660 | Hta3_histone_H2A | 12.75 | Yes |
| TTHERM_00454080 | Rps22_predicted_protein | 9.50 | Yes |
| TTHERM_00498190 | Hhf1_predicted_protein | 9.38 | Yes |
| TTHERM_00125280 | Pdd1_chromodomain_protein | 1.00 | No |

## FIGURES

**Figure 1: Identification and Affinity Purification (AP) of Snf5 and Saf5:** α-FLAG/M2 recognizes the Flag-tag on FZZ in the whole cell extract (WCE) and AP experiments. **A. Expression Analysis/AP of Snf5-FZZ.** Snf5 ~60kDa (18kDa FZZ + 42kDa Snf5). Lanes 1, 3 and 5 are untagged (-) and lanes 2, 4 and 6 are tagged (+, SNF5-FZZ) *Tetrahymena* strains. Lanes 1 and 2 represent WCE prepared with TCA precipitation. Lanes 3-6 were extracted with a soluble-affinity buffer. Some protein degradation is apparent. Snf5 co-purifies with Brg1 (top panel, lane 6) **B. Expression Analysis/AP of Saf5-FZZ.** Saf5 ~83kDa (18kDa FZZ + 65kDa Saf5). Lane 1,2,5 and 6 are untagged (-) and lanes 3, 4, 7, 8 and 9 (positive control) are tagged (+, SAF5-FZZ) *Tetrahymena* strains. Saf5 co-purifies with Brg1 (top panel, lanes 7 and 8). **C. Snf5 and Saf5 localize to the MAC during growth.** Left panels show stained nuclei, macronucleus (mac) and micronucleus (mic), by DAPI. Right panels show localization of the FZZ during vegetative growth.

**Figure 2:** Analysis of *Tetrahymena* BRD-containing proteins. **A: Domain architecture of the identified BRD-containing proteins.** Domains were predicted using the SMART web tool and Pfam domain analysis (see Materials and Methods). **B: Phylogenetic analysis of *Tetrahymena* BRDs.** The amino acid sequences of the predicted BRDs were aligned using MUSCLE. The phylogenetic analysis was carried out using the neighbour joining method with 1000 bootstrap replicas (confidence >90% for all nodes). **C: Predicted structure (left) of the Ibd1 BRD shown in ribbon diagram with rainbow color scheme.** Blue represents the N-terminus whereas red shows the C-terminus of the predicted structure. The superimposition (right) was carried out using the BRD of human SMARCA2 protein (PDB: 5DKC) which is shown in violet color backbone format. Note: The identified *Tetrahymena* BRD-containing proteins were named based on the domain architecture if no clear human ortholog was available. BroW1 stands for bromo-WD40 domain protein; BrEt: Bromo-Et domain protein; BrAn: Bomo-Ank domain protein; Brop1-6: BRD-containing protein.

**Figure 3: Identification and Affinity Purification (AP) of Ibd1. A. Western blot to assess whether *Tetrahymena* transformant cells are expressing Ibd1-FZZ.** Whole cell extract (WCE) using TCA of Ibd1-FZZ cells during vegetative growth. Ibd1 ~50kDa (18kDa FZZ + 32kDa Ibd1). Lane 1 is untagged (-) and lanes 2,3,4,5, and 6 are tagged (+, Ibd1-FZZ) *Tetrahymena* strains **B. Expression analysis of Ibd1-FZZ during vegetative growth.** WCE and AP experiment extracted with a soluble-affinity buffer for untagged (-, lanes 1 and 3) and tagged (+, lanes 2 and 4) Ibd1-FZZ *Tetrahymena* strains. The BDR-containing protein is recognized by α-Flag/M2 and co-purifies with Brg1 (Right top panel, lane 4). **C. Network view of Ibd1 protein-protein interactions.**

The edge thickness represents the averaged spectral counts for the prey. Bait proteins are shown in larger nodes which are colored according to predicted complexes as indicated.

**Figure 4: Ibd1 Expression pattern.** Ibd1-FZZ (B2086) and untagged cells after 24 hours of vegetative growth (VG) and starvation (STV) and after 3, 4.5, 6, and 8 hours post mixing (Mating of Ibd1-FZZ B2086 and CU428) **A. Expression analysis of Ibd1-FZZ during *T. thermophila*'s life cycle.** Whole cell extraction followed by TCA precipitation of untagged and tagged (Ibd1-FZZ) *Tetrahymena* strains. Ibd1 expresses throughout the *T. thermophila* life cycle (bottom panel). Brg1 is a loading control for expression throughout the *T. thermophila* life cycle (top panel). Pdd1 is an exclusively developmental protein and is used as a control during conjugation (middle panel). **B. Ibd1 localizes to the MAC during growth and sexual development including meiosis.** The upper row of each panel shows a cartoon of *T. thermophila* depicting macronucleus (MAC), micronucleus (MIC), gametic nuclei, zygotic nuclei, new MIC, anlagen (new MAC) and old mac (OM) at different stages. Untagged cells are in the second and third panels and tagged (Ibd1-FZZ) are in the 2 lower panels. DAPI localizes to nuclei and Rho α-ZZ to the tagged protein.

**Figure 5: A. Modulation of Ibd1 interactome during conjugation.** Dot plot overview of the interaction partners identified with Ibd1-FZZ during vegetative growth and 5 hours after initiation of conjugation. Inner circle color represents the average spectral count, the circle size maps to the relative prey abundance across all samples shown, and the circle outer edge represents the SAINT FDR. **B. Expression analysis of IBD1-FZZ 5**

**hours into conjugation.** Whole Cell Extract (WCE) and Affinity Purification (AP) of untagged (-, Lanes 1 and 4) and tagged (+, Ibd1-FZZ, lanes 2,3,5, and 6) strains. In the AP samples taken during conjugation Brg1 (Top Panel, Lanes 5 and 6) and Pdd1p cannot be detected (data not shown). Ibd1 is recognized by α-Flag/M2 (Lower Panel, Lanes 5 and 6).

**Figure 6: Ibd1 is localized to actively transcribed genes. A. Ibd1 Occupancy.** Ibd1 shows occupancy for 837 ORF and 396 intergenic regions. When Ibd1 was enriched (IP/INPUT) 2-3 (≥2), 4-5 (≥4), 6-8 (≥6) and more than 8 (≥8) times it was found in 795, 370, 69 and 19 sites respectively. **B. *Tetrahymena*'s expression distribution and Ibd1 localization.** Left panel. 9% of *Tetrahymena*'s genes are highly expressed. Right panel. 457 ORF (54%) that are occupied by Ibd1 are highly expressed. Ibd1 also localizes to 132 (14%) coding regions that do not present available data for the RNA-Seq data (GEO accession GSM692081, [43]). **C. Ibd1 prefers highly expressed genes.** High amounts of Ibd1 occupancy is related to highly expressed genes. The trend shows that the higher the Ibd1 fold enrichment (IP/INPUT) is the higher the occupancy of highly expressed genes. This is evident when Ibd1 is enriched more than 4 times. **D. Ibd1 frequently localizes to housekeeping genes.** The GO terms of the observed genes with significant enrichment (≥4) are genes responsible for cell maintenance. **E. *Tetrahymena*'s GO Terms.** Distribution of *T. thermophila* genes based on GO Biological Functions. **F. ChIP-qPCR validation.** Anti-FLAG ChIP was performed in the 3 replicas of untagged and 4 replicas of Ibd1-FZZ during vegetative growth. ChIP DNA was amplified using primers to amplify *HTA*3, *RPS*22, *HFF*1 and *PDD*1 by real-time PCR using SYBR green. The significant p-values from the t-test are represented by a * (p-value <0.05). These significant p-values are 0.043 for HTA3, 0.041 for RPS22 and 0.015 for HFF1 this confirmed enrichment of Ibd1 in these genes. Our negative control, Pdd1 shows a no significant p-value (ns) meaning no enrichment at this gene. The error bars represent the standard error of the mean for each sample. (see Additional file 9 for Raw Data).

**Figure 7: Ibd1 is localized in promoters, ORF and terminators**. In regions with more than or equal to 4-fold enrichment (IP/NPUT), Ibd1 localizes to 8 specific type-regions, including: **A.** 483 ORF, **B.** 91 promoters, **C.** 33 terminators, **D.** localization in 2 regions where there is overlap between the promoter of one predicted gene and the terminator of another, **E.** 13 regions showed localization to a single 5' promoter region potentially controlling expression of two predicted genes, **F.** localization to 11 single terminators 3' regions of two distinct predicted genes. Combining these data for genes that present enrichment in the ORF and intergenic region, we found that there is mutual enrichment in: **G.** 37 regions that occupy from the promoter to the ORF, and **H.** 19 regions that present enrichment from the ORF to the terminator region, (see Additional file 7 and 8 for Raw Data). The fold enrichments are presented beside each peak.

**Additional files**

Additional_file_1_Snf5_Alignments_Cloning.pptx
Additional_file_2_Saf5_Alignments_Cloning. pptx

Additional_file_3_Ibd1_Alignments_Cloning. pptx

Additional_file_4_Primers.xlsx

Additional_file_5_Peptide_array.xlsx

Additional_file_6_Ibd1_MS_5hConj.xlsx

Additional_file_7_ChIP_seq_ORF.xlsx

Additional_file_8_ChIP_seq_Intergenic.xlsx

Additional_file_9_ChIP_qPCR.xlsx

Additional_file_10_Additional_Methods.docx